\documentclass[9pt,twocolumn,twoside]{osajnl}

\journal{ol} 

\setboolean{shortarticle}{true}

\title{Multipath Wave-Particle Duality in Classical Optics}

\author[1]{Bibhash Paul}
\author[1]{Sammi Kamal}
\author[2,*]{Tabish Qureshi}

\affil[1]{Department of Physics, Jamia Millia Islamia, New Delhi 110025}
\affil[2]{Centre for Theoretical Physics, Jamia Millia Islamia, New Delhi 110025}

\affil[*]{Corresponding author: tabish@ctp-jamia.res.in}

 \doi{\url{http://dx.doi.org/10.1364/OL.392762}}

\begin{abstract}
It is well known that in classical optics, the visibility of interference,
in a two-beam light interference, is related to the optical coherence of
the two beams. A wave-particle duality relation can be derived using this
mutual coherence. The issue of wave-particle duality in classical optics
is analyzed here, in the more general context of multipath interference.
New definitions of interference visibility and path distinguishability 
have been introduced, which lead to a duality relation for multipath
interference. The visibility is shown to be related to a new multi-point
optical coherence function.
\end{abstract}

\setboolean{displaycopyright}{true}

\begin{document}

\maketitle

The dual nature of light has been a subject of debate
for a long time. Corpuscular theory and the wave theory of light constituted
two opposite viewpoints on the nature of light. With the advent of
quantum mechanics, the concept of dual nature of light grew much deeper
as quantum mechanics itself is a wave theory.
Niels Bohr formalized the notion of wave-particle duality by his
principle of complementarity \cite{bohr}. The idea behind it is that
any experiment can fully reveal either the particle nature or the
wave nature of light, or of any other quantum entity, but never both of
them. A two-slit interference experiment, using photons or massive particles,
is an ideal testbed to study the principle of complementarity.
The interference is an obvious signature of the wave nature in such an
experiment. Finding out which of the two paths the particle followed, 
amounts to revealing its particle nature. Wooters and Zurek \cite{wootters}
tried to investigate Bohr's principle by asking if one can
{\em partially} observe both wave and particle natures at the same time?
The expectation was that if an experiment only partially reveals (say) the
particle nature, it may also allow partial revelation of the wave nature.
The sharpness of interference may constitute a measure of wave nature,
and the amount of knowledge, regarding which slit the particle went through,
may quantify the particle nature. Wooters and Zurek's line of investigation
was later extended by Englert who derived a wave-particle duality
relation \cite{englert}
\begin{equation}
{\mathcal D}^2 + {\mathcal V}^2 \le 1,
\label{englert}
\end{equation}
where ${\mathcal D}$ is path distinguishability, a measure of the particle
nature, and ${\mathcal V}$ the visibility of interference, a measure of
wave nature. The above study assumed that a measuring device is introduced
which can tell which of the two slits the particle went through.

\begin{figure}
\centerline{\resizebox{9.0cm}{!}{\includegraphics{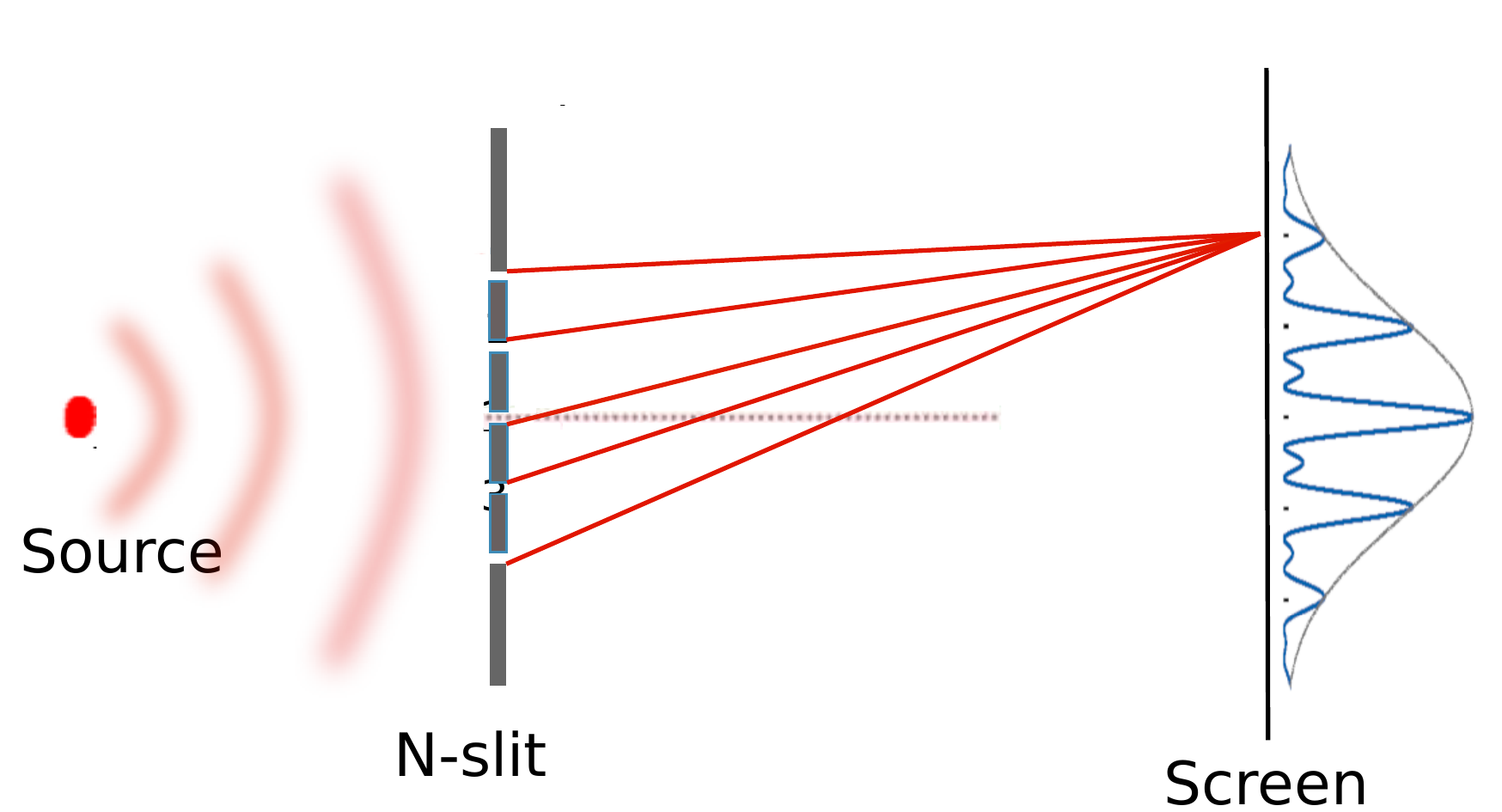}}}
\caption{A schematic representation of multislit interference experiment.
 Any point on the screen recieves contributions from all slits.
}
\label{nslit}
\end{figure}

However, there is another line of thought in which wave-particle duality
can be studied, without introducing a which-way measuring,
or path-detecting device. If the two beams emerging from the two slits,
have different intensities, one could \emph{predict} which of the two paths the 
particle might have taken, with a success which is more than a random
50-50 guess.  Greenberger and Yasin \cite{greenberger}, and later
Jaeger, Shimoni and Vaidmann \cite{jaeger}, followed this line of thought
and derived different kind of duality relation
\begin{equation}
\mathcal{P}^2 + \mathcal{V}^2 \le 1,
\label{GY}
\end{equation}
where $\mathcal{P}$ is a path-predictability, and $\mathcal{V}$ the
visibility of interference. The ability to predict the path of the particle,
allows one to argue that there is a certain degree of particle nature
associated with the objects passing through the double-slit. The fringe
visibility, which quantifies the wave nature, is taken to be just the
Michelson's fringe contrast \cite{bornw}
\begin{equation}
{\mathcal V} = \frac{I_{{max}} - I_{{min}}}{ I_{{max}} + I_{{min}} } ,
\label{V}
\end{equation}
where the notations have the usual meaning.
Although both inequalities (\ref{englert}) and (\ref{GY})
have been referred to as wave-particle duality relations in the literature, 
it is important to remember that the two are different, and pertain to
different experimental situations. However, the name distinguishablity
is being used interchangeably for both $\mathcal{D}$ and $\mathcal{P}$.

The issue of wave-particle duality, in the realm of classical optics,
has come into a lot of recent attention
\cite{eberly1,eberly2,eberly3,copt1,copt2,copt3}.
It should be emphasized here that for classical light, since one does not talk
at the level of single photons, it is meaningless to talk about duality
relations of the type (\ref{englert}). One can only predict with a good
success rate that light would pass through the more intense beam. Then it is
meaningful to talk of duality relation of the kind (\ref{GY}). 

Let us now consider a scenario where light passes through an array of
equally spaced $n$ slits (see Fig. \ref{nslit}). The complex amplitude of the light field
depends on its two-dimensional transverse coordinate $r$, time $t$,
and intrinsic polarization $\hat{s}$. The field at the location of
the $i$'th slit, at time $t$, has arbitrary amplitude and unit polarization:
$E_i = \hat{s}_i E_i(r_i,t)$. The field recieved from the $i$'th slit,
at a point $r_x$ on the detection screen, is given by 
\begin{equation}
 E_{ix} = K_i \hat{s}_i E_i(r_i,t-t_i),
\end{equation}
where $t_i$ is the time taken by light to travel from the $i$'th slit to
the point $r_x$ on the screen, and $K_i$ is the usual purely imaginary
propagation factor \cite{bornw}. The total intensity at the point 
$r_x$ on the screen is a result of fields received from all the $n$ slits:
\begin{eqnarray}
I(r_x,t) &=& \left| \sum_{i=1}^n K_i \hat{s}_i E_i(r_i,t-t_i)\right|^2
\nonumber\\
&=& \sum_{i=1}^n |K_i|^2 \left| E_i(r_i,t-t_i)\right|^2 \nonumber\\
&&+ \sum_{i\ne j} K_iK_j^* \langle E_i(r_i,t-t_i)E_j^*(r_j,t-t_j)\rangle
\langle\hat{s}_i\cdot\hat{s}_j\rangle , 
\end{eqnarray}
where the angular brackets denote statistical average.
The field correlation functions can be translated in time. With this in mind,
the above relation can be written as
\begin{eqnarray}
I(r_x,t) &=& \sum_{i=1}^n I_i 
+ \sum_{i\ne j} |K_i||K_j| \langle\hat{s}_i\cdot\hat{s}_j\rangle
 \Re \Gamma_{ij}(r_i,r_j,\tau_{ij}) ,
\label{I1}
\end{eqnarray}
where $\tau_{ij}=t_i-t_j$, and $\Gamma_{ij}(r_i,r_j,\tau_{ij})$ is the
vector-mode mutual coherence function between the $i$'th and $j$'th slits,
given by \cite{eberly3}
\begin{equation}
\Gamma_{ij}(r_i,r_j,\tau) = \langle\hat{s}_i\cdot\hat{s}_j\rangle
\langle E_i(r_i,0)E_j^*(r_j,\tau)\rangle .
\end{equation}
Here $I_i$ is to be interpreted as the intensity at the point $r_x$
on the screen, if only the $i$'th slit were open.
The mutual coherence function can be normalized as
\begin{equation}
\gamma_{ij}(\tau) = \frac{\Gamma_{ij}(r_i,r_j,\tau)}
{\sqrt{\Gamma_{ii}(r_i,r_i,0)\Gamma_{jj}(r_j,r_j,0)}}
\end{equation}
For monochromatic light $\gamma_{ij}(\tau)$ may be represented as \cite{bornw}
$\gamma_{ij}(\tau) = |\gamma_{ij}(\tau)|e^{i\omega\tau+i\phi_{ij}}$,
where $\phi_{ij}$ are certain unspecified phases. Using this,
eqn. (\ref{I1}) assumes the following form
\begin{eqnarray}
I(r_x,t) &=& \sum_{i=1}^n I_i 
+ \sum_{i\ne j} \sqrt{I_i I_j}
 |\gamma_{ij}(\tau_{ij})| \cos(\omega\tau_{ij}+\phi_{ij}) ,
\label{I2}
\end{eqnarray}
where $\phi_{ij}$ are certain phase factors.
The above relation describes the intensity at a point on the screen,
in terms of a sum of normalized mutual coherences between all pairs of
slits. 

\begin{figure}
\centerline{\resizebox{9.0cm}{!}{\includegraphics{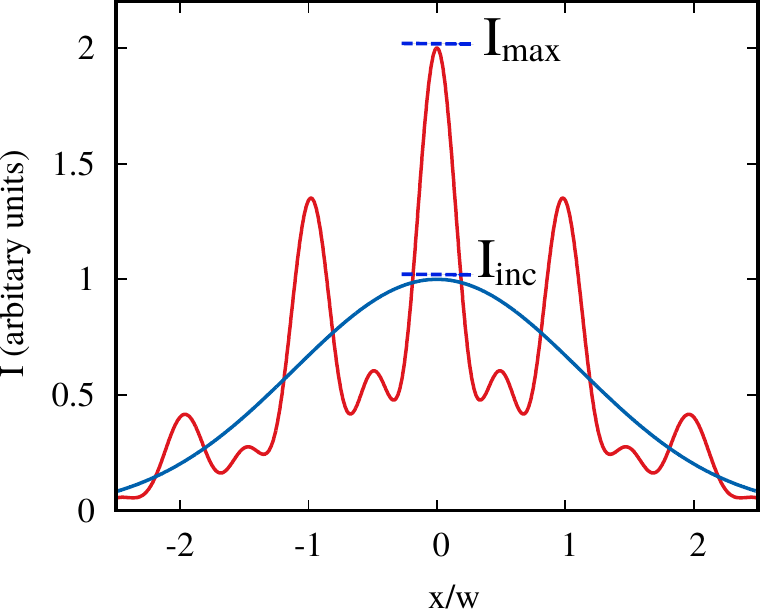}}}
\caption{A representative multislit interference pattern is shown here,
as a function of the position on the screen $x$, scaled by the primary
fringe width $w$.
When incoherence leads to a loss of interference, 
the broad Gaussian profile (shown in the figure) is what remains.
 $I_{max}$ and $I_{inc}$ are indicated in the figure.
}
\label{interf}
\end{figure}

Now if one tries to evaluate the visibility of interference from (\ref{I2})
using (\ref{V}), one does not get a compact and elegant answer. However, one
can check that for $n=2$, one does recover the known result
$\mathcal{V}=|\gamma_{12}(\tau_{12})|$, which indicates that we are on the
right track. For $n > 2$ one has to look for another strategy.
Looking at (\ref{I2}) one would notice the first term represent the incoherent
contribution to the interference, basically the sum of intensities reaching
a point on the screen from each slit separately, without any interference.
The second term represents interference. If one subtracts out the 
incoherent contribution from the full interference pattern, what remains
is just the interference term, i.e., the second term. Following this line
of thought, a new definition of visibility has been introduced 
\cite{tania,cohrev}
\begin{equation}
\mathcal{V}_{C} = \frac{1}{n-1} \frac{I_{{max}} - I_{{inc}}}{ I_{{inc}} } ,
\label{Vn}
\end{equation}
where $n$ is the number of slits, $I_{max}$ is the intensity at a primary
maximum of the interference pattern, and $I_{{inc}}$ is the intensity
at the position of a primary maximum if the light is made incoherent
before entering the slits, and consequently {\em the interference is destroyed}
(see Fig. \ref{interf}).
For example, in (\ref{I2}), the first term represents $I_{{inc}}$.
For simplicity, let us assume that all $\phi_{ij}$ are zero. 
In (\ref{I2}), the maximum intensity will be when all cosine terms are
equal to unity at the same time. These are the locations of primary maxima.
Using (\ref{I2}) and (\ref{Vn}) leads one to
\begin{eqnarray}
\mathcal{V}_{C} &=& \frac{1}{n-1}
\sum_{i\ne j}  \frac{ \sqrt{I_i I_j}}{\sum_{k=1}^n I_k }
|\gamma_{ij}(\tau_{ij})| .
\label{Vcvalue}
\end{eqnarray}
First thing to notice is that for $n=2$,
$\mathcal{V}_{C}  = |\gamma_{12}(\tau_{12})|$, which shows that for $n=2$,
$\mathcal{V}_{C}$ gives the same value as the traditional visibility (\ref{V}).
If intensities at all the slits are equal, then
\begin{eqnarray}
\mathcal{V}_{C} &=& \frac{1}{n(n-1)}
\sum_{i\ne j}  |\gamma_{ij}(\tau_{ij})| ,
\label{Vgamma}
\end{eqnarray}
which is just the average of normalized mutual coherence over all pairs
of slits. It is satisfying to see that, even for multislit interference,
interference visibility is related to the degree of mutual coherence. 

Next we turn our attention to the degree of distinguishability of paths,
which is technically predictability, but in optics literature it is 
called distinguishability. We will also refer to it here as distinguishability,
but keeping in mind that it only pertains to guessing the path when the
paths are unequal. For the case of two slits, it is defined as
\begin{equation}
{\mathcal D} = \frac{|I_1 - I_2|}{ I_1 + I_2 } ,
\label{D2}
\end{equation}
which means that if the intensities on the screen, coming from the two slits,
are equal, there is no way one can distinguish between the two paths,
and $\mathcal{D} = 0$.
On the other extreme, if the intensity (say) $I_2$ is negligibly small,
one can be almost sure that any light reaching the screen came from slit 1,
and $\mathcal{D} \approx 1$.
Now if one moves on to n-slit interference, there is no obvious way to
generalize (\ref{D2}) to more than two slits.
We start by rewriting (\ref{D2}) as
\begin{equation}
\mathcal{D} = \sqrt{1 - \frac{ 4I_1 I_2 }{(I_1 + I_2)^2}} .
\label{PGY}
\end{equation}
Note that the second term in the square root in the above equation
can also be written as $(\sum_{j\neq i} \sqrt{I_i I_j}/(I_1 + I_2))^2$. 
This form can be intuitively generalized to n-slit case.
We introduce the following path distinguishability for the case of $n$ slits, 
\begin{equation}
\mathcal{D} \equiv \sqrt{1 - \left({1\over n-1}\sum_{i\neq j}
\frac{\sqrt{I_i I_j}}{\sum_{k=1}^n I_k } \right)^2},
\label{Dn}
\end{equation}
where the role of the factor $\frac{1}{n-1}$ is to normalize the
second term inside the square root. For $n=2$, (\ref{Dn}) reduces to
(\ref{PGY}).

Before proceeding further, it may be worthwhile to test this new
disntinguishability in various limits.
If the intensities coming from all the slits, except one, are zero,
we should have complete knowledge about which slit the light comes from.
As expected, in such a situation, (\ref{Dn}) yields $\mathcal{D} = 1$.
If the intensities from all the slits are equal (i.e., all $I_{j}=I_0$),
there is no way one can distinguish between the $n$ paths. For this case
(\ref{Dn}) yields $\mathcal{D} = 0$.

Now that we have our $n$-path distinguishability and generalized visibility, 
defined, we proceed to finding out the constraints on their values together.
Using (\ref{Vcvalue}) and (\ref{Dn}), we can write
\begin{eqnarray}
\mathcal{D}^2 + \mathcal{V}_C^2 &=& 1 - \left({1\over n-1}\sum_{i\neq j}
\frac{\sqrt{I_i I_j}}{\sum_{k=1}^n I_k } \right)^2 \nonumber\\
&& + \left(\frac{1}{n-1}
\sum_{i\ne j}  \frac{ \sqrt{I_i I_j}}{\sum_{k=1}^n I_k }
|\gamma_{ij}(\tau_{ij})|\right)^2
\label{DV0}
\end{eqnarray}
Since $|\gamma_{ij}| \le 1$, we can write the inequality
\begin{eqnarray}
\mathcal{D}^2 + \mathcal{V}_C^2 ~ \le ~ 1 .
\label{DV}
\end{eqnarray}
This is a new wave-particle duality relation involving the path
distingushability and interference visibility, and works for interference
from any number of slit. The inequality saturates if 
all $|\gamma_{ij}| = 1$ (fully coherent light). For $n=2$ the inequality
reduces to the known duality relation for two-slit interference.

In the following we would like to propose another way of defining the 
degree of distinguishability of paths, which will lead to a simpler
duality relation. Instead of (\ref{Dn}) we propose the following
distinguishability
\begin{equation}
\mathcal{D}' \equiv 1 - {1\over n-1}\sum_{i\neq j}
\frac{\sqrt{I_i I_j}}{\sum_{k=1}^n I_k } .
\label{Dnp}
\end{equation}
This definition is simpler in form, without involving the square of a
sum. For $n=2$, it reduces to
\begin{equation}
\mathcal{D}' = 1 - \frac{2\sqrt{I_1 I_2}}{I_1 + I_2 } .
\label{D2p}
\end{equation}
which can also be written as
\begin{equation}
\mathcal{D}' = \frac{(\sqrt{I_1} - \sqrt{I_2})^2}{ I_1 + I_2 } .
\label{D2pn}
\end{equation}
As one can see, this is not any more complicated than (\ref{D2}), and
follows the same physical idea.

Now it is easy to see that (\ref{Dnp}), together with (\ref{Vcvalue}),
yields
\begin{eqnarray}
\mathcal{D}' + \mathcal{V}_C &=& 1 - \frac{1}{n-1} \sum_{i\ne j}
 \frac{ \sqrt{I_i I_j}}{\sum_{k=1}^n I_k } (1-|\gamma_{ij}(\tau_{ij})|) .
\label{DVp0}
\end{eqnarray}
Since $|\gamma_{ij}| \le 1$, we can write a new duality relation
\begin{eqnarray}
\mathcal{D}' + \mathcal{V}_C ~ \le ~ 1 .
\label{DVp}
\end{eqnarray}
For the case of two slits, this reduces to the following duality relation
for two-slit interference
\begin{eqnarray}
\mathcal{D}' + \mathcal{V} ~ \le ~ 1 ,
\label{DV2}
\end{eqnarray}
where $\mathcal{V}$ is the traditional visibility (\ref{V}).
As one can see, this is a simpler duality relation and using the squares
of the two measures was an unnecessary baggage, being carried along just because
of trying to mimic the form of the first duality relation (\ref{GY})
derived by Greenberger and Yasin \cite{greenberger}.

Finally we take a closer look at (\ref{Vgamma}), where the new visibility
$\mathcal{V}_C$ seems to suggest a n-point degree of coherence. We suggest
a general n-point degree of mutual coherence as:
\begin{equation}
\gamma_n(r_1,r_2,\dots ,r_n,\tau_{12},\tau_{23},\dots) =
\frac{1}{n(n-1)} \sum_{i\ne j}  |\gamma_{ij}(r_i,r_j,\tau_{ij})| ,
\label{gamman}
\end{equation}
which is a real quantity,  with magnitude $0 \le \gamma_n \le 1$.
It appears to be the classical optical counterpart of a recently introduced
measure of quantum coherence $\mathcal{C}$ \cite{coherence,cd}:
\begin{equation}
\mathcal{C} =
\frac{1}{n-1} \sum_{i\ne j}  |\rho_{ij}| ,
\label{C}
\end{equation}
where $|\rho_{ij}|$ are the matrix elements of the density operator of
the particle, in a particular basis. We claim 
so because in multipath \emph{quantum} interference, the new way of measuring
visibility, that is described here, yields the quantum coherence 
$\mathcal{C}$, where the basis is fixed by the $n$ paths
of the particle through the slits \cite{tania,cohrev}. 
So, for n-path interference, the new visibility $\mathcal{V}_C$ is given by
the n-point degree of coherence $\gamma_n$ in classical optics, and by
the quantum coherence $\mathcal{C}$ in quantum mechanics.
Since this n-point degree of coherence, at the multislit,
is directly related to the visibility of interference, we believe it may turn
out to be useful in some other situations too. 
In general, the quantum coherence $\mathcal{C}$ is basis dependent, which is
sometimes considered its weakness. On the other hand, the n-point
degree of coherence $\gamma_n$ is not dependent on any `basis'.
Given the $n$ points for which the coherence is sought, there
is only one way in which $\gamma_n$ is to be calculated.
In this sense it appears to have a more universal significance.
The usefulness of 
$\gamma_n(r_1,r_2,\dots ,r_n,\tau_{12},\tau_{23},\dots)$ needs to be
explored further.

In summary, wave-particle duality in multipath interference is now well
studied in the quantum domain \cite{cd,3slit,nslit,bagan,coles1,predict},
however, it has not been well studied in classical optics.
We have analyzed multipath interference in classical optical
domain, and introduced a new way of measuring interference visibility.
This visibility turns out to be related to a new multipoint 
degree of mutual coherence. We
also introduced a quantitative measure of the degree of distinguishability
of the paths in a n-beam interference. These two together, lead to a new
wave-particle duality relation. In another extension of the analysis we
define yet another path distinguishability, which leads to a simpler
wave-particle duality relation. For two-slit interference, it leads to
a new duality relation which is simpler than the one currently known.
The new multipoint degree of mutual coherence appears to the 
classical optical analogue of a recently introduced measure of quantum
coherence, in the context of quantum information theory. Further study is
needed to explore the usefulness of the multipoint mutual coherence.

\medskip

\noindent\textbf{Acknowledgement.} Bibhash Paul and Sammi Kamal thank the Centre for Theoretical Physics
at Jamia Millia Islamia, for providing the facilties of the Centre during
the course of this work.

\medskip

\noindent\textbf{Disclosures.} The authors declare no conflicts of interest.


\begin{thebibliography}{0}
\bibitem{bohr} N. Bohr, ``The quantum postulate and the recent development of
atomic theory," {\em Nature (London)} {\bf 121}, 580-591 (1928). 

\bibitem{wootters} W. K. Wootters and W. H. Zurek,
``Complementarity in the double-slit experiment: Quantum nonseparability
and a quantitative statement of Bohr's principle",
{\em Phys. Rev. D} {\bf 19}, 473 (1979).

\bibitem{englert} B-G. Englert, ``Fringe visibility and which-way information:
an inequality", {\em Phys. Rev. Lett.} {\bf 77}, 2154 (1996).

\bibitem{greenberger} D.M. Greenberger, A. Yasin,
``Simultaneous wave and particle knowledge in a neutron interferometer",
Phys. Lett. A 128, 391 (1988).

\bibitem{jaeger} G. Jaeger, A. Shimony, L. Vaidman, ``Two interferometric complemenarities," {\em Phys. Rev. A} {\bf 51}, 54 (1995).

\bibitem{bornw} Born M and Wolf E Principles of Optics, 3rd ed. (Pergamon, New York, 1965).

\bibitem{eberly1} X.-F. Qian, T. Malhotra, A. N. Vamivakas, J. H. Eberly, “Coherence constraints and the last hidden optical coherence,” Phys. Rev. Lett. 117, 153901 (2016).

\bibitem{eberly2} J. H. Eberly, X.-F. Qian, and A. N. Vamivakas, “Polarization coherence theorem,” Optica 4, 1113 (2017).

\bibitem{eberly3} X.-F. Qian, A. N. Vamivakas, J. H. Eberly, “Entanglement limits duality and vice versa,” Optica 5, 942 (2018).

\bibitem{copt1} F. De Zela, “Optical approach to concurrence and polarization,” Opt. Lett. 43, 2603 (2018).

\bibitem{copt2} B. Kanseri, Sethuraj K. R., “Experimental observation of the polarization coherence theorem,” Opt. Lett. 44, 159 (2019).

\bibitem{copt3} A. Norrman, A. T. Friberg, G. Leuchs, “Vector-light quantum complementarity and the degree of polarization ,” Optica 7, 93 (2020).

\bibitem{tania} T. Paul, T. Qureshi, ``Measuring quantum coherence in multi-slit interference,"
 {\em Phys. Rev. A} {\bf 95}, 042110 (2017).

\bibitem{cohrev} T. Qureshi, ``Coherence, interference and visibility," \href{http://dx.doi.org/10.12743/quanta.v8i1.87}{{\em Quanta} \textbf{8}, 24 (2019)}.

\bibitem{coherence} T. Baumgratz, M. Cramer, M.B. Plenio,
``Quantifying Coherence",
{\em Phys. Rev. Lett.} {\bf 113}, 140401 (2014).

\bibitem{cd}  M.N. Bera, T. Qureshi, M.A. Siddiqui, A.K. Pati,
``Duality of quantum coherence and path distinguishability", {\em Phys. Rev. A} {\bf 92}, 012118 (2015).

\bibitem{3slit}  M.A. Siddiqui, T. Qureshi, ``Three-slit interference:
A duality relation", {\em Prog. Theor. Exp. Phys.} {\bf 2015}, 083A02 (2015).

\bibitem{nslit}  T. Qureshi, M.A. Siddiqui, ``Wave-particle duality in n-path interference", {\em Ann. Phys.} {\bf 385}, 598-604 (2017).

\bibitem{bagan} E. Bagan, J. Calsamiglia, J.A. Bergou, M. Hillery, ``Duality Games and operational duality relations," {\em Phys. Rev. Lett.} {\bf 120}, 050402 (2018).

\bibitem{coles1}  P.J. Coles, ``Entropic framework for wave-particle duality in multipath interferometers," {\em Phys. Rev. A} {\bf 93}, 062111 (2016).

\bibitem{predict}  P. Roy, T. Qureshi, ``Path predictability and quantum coherence in multi-slit interference," {\em Phys. Scr.} 94, 095004 (2019).


\end{thebibliography}
\end{document}